\documentstyle[12pt,psfig]{article}
\def\reference{\parskip 0pt\par\noindent\hangindent 0.5 truecm}

\begin{document}
%
%
\title{The Size of IDV Jet Cores}
%


\author{T. Beckert, 
        T. P. Krichbaum, G. Cim\`o, L. Fuhrmann, A. Kraus, \\ A. Witzel,
        J. A. Zensus 
} 

\date{}
\maketitle

{\center
     Max-Planck-Institut f\"ur Radioastronomie, \\ Auf dem H\"ugel 69,
     53121 Bonn, Germany \\ tbeckert@mpifr-bonn.mpg.de \\[3mm]
}

%
\begin{abstract}
Radio variability on timescales from a few hours to several days in
extragalactic flat-spectrum radio sources is generally classified as
intra-day variability (IDV). The origin of this short term variability is
still controversial and both extrinsic and intrinsic mechanisms must be
considered and may both contribute to the observed variations.
 The measured linear and circular polarization of IDV
sources constrains the low energy end of the electron population. 
Any population of cold electrons within sources at or above the equipartition
temperature of $10^{11}$ K depolarizes the emission and can be ruled out.
Intrinsic shock models are shown to either violate the large fraction
of sources displaying IDV or they do not relax the light travel time argument
for intrinsic variations. From structure function analysis, we further
conclude that interstellar scintillation also leads to tight size estimates
unless a very local cloud in the ISM is responsible for IDV.
\end{abstract}


\bigskip

%
%

\section{Introduction}
Intraday Variability (IDV) of flat-spectrum compact Quasar cores and BL Lacs
at cm-wavelength has been discovered in 1985 (Heeschen et al., 1987)
and is a common phenomenon ($\sim 25\%$) (Quirrenbach et al., 1992) among
these sources.
On rare occasions in 0716+714 and 0954+658, correlations with optical
variations have been found and the evidence has been reviewed several times
(e.g. Wagner S.J., Witzel A., 1995; Krichbaum, this volume). Radio-optical
correlations, if real,  suggest either fast intrinsic variations 
or gravitational lensing. From the light travel time argument the 
apparent brightness temperatures $T_b$ are in the range of 
$10^{17}$--$ 10^{21}$K and far in excess of the intrinsic inverse 
Compton (IC) limit of $10^{12}$K.
The observed superluminal motions of jet components in many of these
sources imply $\Gamma=5$--$10$ and allow for Doppler boosting factors
${\cal{D}} =\left(\Gamma (1-\beta \cos\theta)\right)^{-1}$
and time shortening,
which are insufficient for reducing the apparent $T_b$ down to the
IC limit. The required Doppler factors $\cal{D}$ are in the range of
60 up to 1000. Furthermore jets with a surface brightness at the
IC limit are radiatively inefficient (Begelman, Rees \& Sikora, 1994) and
carry most of their energy as bulk motion. This is not supported
by the observed power of radio lobes
of these sources and raises the
energy requirement for the central engine to an uncomfortable level. It is
therefore argued (Readhead, 1994; Begelman, this volume) that incoherent
synchrotron sources in jets should radiate at the equipartition temperature
$T_E \approx 10^{11}$\,K. This enforces Doppler factors which are 2--3 times
larger than for the IC limit. Furthermore synchrotron sources at the IC limit
with bulk motions of $\Gamma > 100$ are
dominated by inverse Compton scattering of either AGN photons
(Begelman, Rees \& Sikora, 1994) or CMB photons
at redshifts $z \sim 1$ and not by the SSC
process. The cooling is catastrophic, independent of the brightness
temperature, and this explanation must therefore be discarded.

Other more tempting suggestions are the propagation of relativistic
thin shocks in the jets (Qian et al., 1991), so that the observed variability
timescale is not a measure of the source size, and scintillation induced
by the interstellar medium (e.g. Rickett 1990) of otherwise
non-variable sources. We will explore both explanations in the following
sections.

\section{Cool Particle Depolarization}
The emission of incoherent synchrotron sources with high brightness
temperatures is dominated by radiation from the $\tau=1$ surface.
For intrinsic $T_b$'s above the equipartition temperature, the energy
of the source is dominated by particles ($e^-,e^+,p$) with a strong
dependence on $T_b$:
\[ U_e/U_B \propto T_b^{7+2\alpha}\quad , \]
where $\alpha$ is the optically thin spectral index of the emission.
Above the IC limit any contribution of cold electrons in an $e^-/p$ plasma
will depolarize the synchrotron emission. Emission from the $\tau=1$ surfaces
of flat spectrum radio cores comes predominantly from electrons with constant
$\gamma_{\mathrm{rad}}$-factor, independent of frequency. We consider
power-law distributions 
of electrons $N(\gamma) \propto \gamma^{-p}$ above a lower cut-off
$\gamma_{\mathrm{min}}$. The
degree of linear polarization is likely to be reduced by tangled magnetic
fields in a turbulent plasma with a typical wavenumber $k_0$ in sources
of size $R$ (e.g. jet radius). For independent variations of magnetic field
orientation in adjacent cells of size $k_0^{-1}$ 
the fractional linear polarization
is $ \pi_L \propto (k_0 R)^{-3/2}$. Further depolarization inside one cell
occurs, if Faraday rotation by electrons around $\gamma_{\mathrm{min}}$
(Jones \& O'Dell 1977)
depolarizes the radiation within the cell $ \pi_L \propto  \tau/\tau_F$.
Here $\tau_F$ is
the Faraday depth in the cell, which is smaller than the Faraday depth for
the whole source by a factor $(k_0 R)^{-1}$. The fractional linear
polarization can then be approximated by
\begin{equation}\label{depol}
  \pi_L \approx \frac{\alpha+1}{\alpha+5/3}\; (k_0 R)^{-1/2}\;
  (\gamma_{\mathrm{rad}}/\gamma_{\mathrm{min}})^{-p}\;
  \frac{\gamma_{\mathrm{min}}}{\ln \gamma_{\mathrm{min}}} \quad .
\end{equation}
For sources above the IC limit the magnetic field strength must be very low
and radiation at GHz-frequencies is emitted by electrons with large
$\gamma_{\mathrm{rad}}$.
The resulting ratio $\gamma_{\mathrm{rad}}/\gamma_{\mathrm{min}}$
in Eq.\ref{depol} becomes large, and
the polarization drops below 1\% if $\gamma_{\mathrm{min}} < 10^3$ at
$T_b = 10^{12}$\,K.
At the equipartition temperature we still have depolarization
if $\gamma_{\mathrm{min}} < 80$. The fact that most IDV sources are variable
in polarized flux with a mean $\pi_L \ge 1\%$ requires a cut-off in the
electron population close to $\gamma$, which dominates the radiation
from $\tau=1$ surfaces. At $T_b > 10^{13}$K this requires fine-tuning of
$\gamma_{\mathrm{min}}$ and provides strong constrains for acceleration
mechanisms. Any substantial population of cold electrons with
$\gamma_{\mathrm{min}} \approx 1$ is
excluded in polarized sources with $T_b > 10^{10}$\,K. This problem does not
arise in pair plasma jets, because Faraday depolarization does not occur there.

\section{The alignment problem of thin shock propagation}
The thickness $\Delta z$ of a layer of post shock gas behind a
relativistic shock, which travels with a shock-Lorenz-factor
$\Gamma_S > 2$ for a distance $z$ is
\begin{equation}
  \Delta z = \left( 1- \sqrt{\frac{\Gamma_S^2 - 4}{\Gamma_S^2 -1}}\right) z
  \quad ,
\end{equation}
assuming that the gas leaves the shock at the sound speed and has
an ultra-relativistic equation of state. A gas element that passed through
the shock at $z=0$ 
is separated by a distance $\Delta z$ from the shock, when the shock has
travelled a distance $z$ in the jet frame. For oblique shocks (as suggested by
Spada et al., 1999) the velocity and the thickness of the
post shock gas will be larger. The ratio
of thickness to the square-root of the surface area is
\begin{equation}
  \tilde{\tau} \approx \Delta z/(\sqrt{\pi} z \sin \psi) =
  \left( 1- \sqrt{\frac{\Gamma_S^2 - 4}{\Gamma_S^2 -1}}\right)
  /(\sqrt{\pi} \sin \psi) \quad ,
\end{equation}
where it is assumed that the shock travelled at a constant velocity from
the tip of a conical jet with half opening angle $\psi$. In the following we
will assume that the shock travels only $1/10$ of that distance.
When the shock is viewed face on,
the surface area is a factor $1/\tilde{\tau}^2$ larger than inferred from the
variability timescale. But this factor is subject to relativistic aberration
\begin{equation}
  \tau = {\cal{D}}\sin\theta +
  \Gamma\,{\cal{D}}\,\tilde{\tau}\,(\cos\theta -\beta) \quad ,
\end{equation}
where $\beta$ is the velocity of the post-shock gas 
and $\theta$ the angle between jet and the line of sight.
The true observed flux variations can arise from orientation changes of
shock and jet
\begin{equation}
  \frac{\Delta F_{\mathrm obs}}{F_{\mathrm int}} = f_1 =
  3 {\cal{D}}^2\,\frac{\partial {\cal{D}}}{\partial \theta}
              \tau^{-2} \Delta\theta  \quad ,
\end{equation}
or from inhomogeneities in particle density or magnetic field
along the jet (Qian et al., 1991),
which are scanned by the shock and travel with the sound speed through
the post-shock gas
\begin{equation}
  \frac{\Delta F_{\mathrm obs}}{F_{\mathrm int}} = f_2
   = {\cal{D}}^3 (\tau/0.667)^{-2}  \quad .  
\end{equation}
 \begin{figure}[t]
 \begin{center}
 \psfig{file=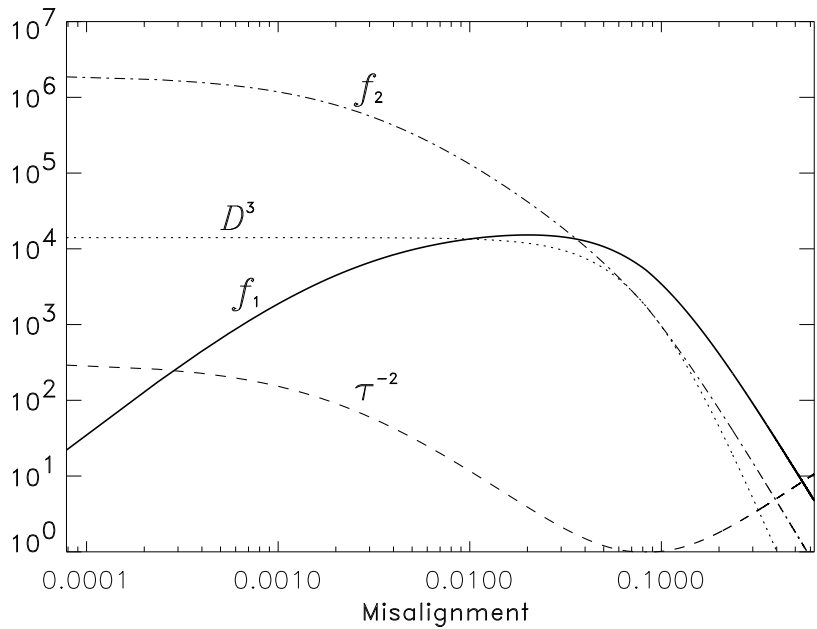,height=9cm}
 \caption{The square of the observable reduced timescale
   $\tau^{-2}$ (dashed) and the
   Doppler boost ${\cal{D}}^3$ (dotted) for the brightness temperature
   in case of intrinsic variations are shown together with the
   {\it flux-boosting factors}
   $f_1$ (solid) and $f_2$ (dash-dotted) as defined in the text. The
   misalignment is the angle $\theta$ between jet direction and line of
   sight. The assumed shock is travelling with $\Gamma_S =6$ in a jet of
   $\Gamma_{\mathrm pre} =2.7$. The implied increase of observed $T_b$ due
   to orientation changes $f_1$ is not substantially larger than
    ${\cal{D}}^3$ and
   the result of scanned perturbations $f_2$ is only sufficient for explaining
   IDV for strong alignment of jet and observer.}
 \label{No_Shock}            
 \end{center}
 \end{figure}
In Fig.1 the Doppler boosting of the flux due to intrinsic variations
for a spherical source ${\cal{D}}^3$ is compared to the {\it
flux-boosting factors}
$f_1$ and $f_2$ for
a shock with $\Gamma_S =6$ in a jet with $\psi = 2.5^\circ$ and
$\Gamma_{\mathrm pre}=2.7$ for the preshocked gas (resulting in a
post shock gas with $\Gamma = 12$). Orientation changes
of thin shocks require
$\Delta \theta \sim \theta$ and consequently $f_1$ can not substantially
increase the flux variations on short timescales compared to
variability in a spherical source. Only if disturbances
along the jet are highlighted by the passage of a shock can the flux variations
be so rapid and strong that $f_2$ exceeds $10^6$ which is necessary to
bring a source with intrinsic $T_b \le 10^{12}$\,K up to the observed
$10^{18}$\,K. Nonetheless this cannot provide an explanation for most
IDV sources, because it requires a misalignment of the line of sight
to the jet direction
much smaller than $1/(3 \Gamma)$, where the plateau of ${\cal{D}}^3$ starts
(see Fig.1). In the specific case shown in Fig.1 the misalignment must be less
than $10^{-3} \Gamma^{-1}$, which cannot be reconciled with  25\%
of core-dominated sources showing IDV.
Furthermore scanned perturbations in the jet lead to variable
optical depth in the shocked gas and timelags, which are not observed
in cross correlations of IDV at different radio frequencies.

\section{Refractive Interstellar Scintillation}
It is known from pulsar measurements that compact radio sources are subject
to scintillation in the ISM of our galaxy. Extragalactic sources 
flicker (Heeschen 1984) at frequencies $\nu \approx 1$\,GHz with timescales of
several days to weeks and this has been interpreted as a result of strong
refractive scattering in the extended ISM of the galactic disk
(Blandford, Narayan \& Romani, 1986).
The transition from strong to weak scattering is expected at
about 5\,GHz depending on the path length in the disk towards the source. The
maximum of the modulation index $m$ is $\approx 1$ for a point source.
A 1Jy source with an apparent $T_b$ of $10^{12}$\,K has a size of $230\,\mu$as
at $\lambda 6$cm and scintillation will be quenched\footnote{A gaussian
brightness distribution is assume for extended sources in this paper.},
because the source size
is much larger than the angular size of the Fresnel scale in the ISM.
For a typical distance of 200\,pc to the scattering medium, the expected
modulation index is $m \le 1.7$\%, while a source with $T_b = 10^{13}$\,K
will have $m \le 5 $\% with a characteristic timescale of 0.5 days.
Therefore any 1Jy source with apparent $T_b$ above few $\times 10^{12}$\,K is
expected to scintillate with a timescale 
comparable to those observed in IDV sources, as has been demonstrated
by Rickett et al. (1995) for 0917+624.

The light curves of scintillating sources contain further information,
which can be explored via structure functions (SF)
(Simonetti, Cordes \& Heeschen, 1985):
\begin{equation}
   \mathrm{SF}(\tau) = \left< [I(t+\tau)-I(t)]^2\right>_t   \quad .
\end{equation}
The SF extracted from the data can be compared to theoretical expectations.
For strong refractive and weak scintillation in an extended medium
Coles et al. (1987) gave a closed form for SF's, which includes
three cut-offs for the
Fourier transform of the local SF. These are due to the Fresnel scale,
the visibility amplitude, which contains the size of the source, and
an exponential cut-off in strong scintillation giving rise to the
refractive scale. Whichever comes first determines the modulation index
and the slope of the SF below the first maximum (see Fig.2).
Following Blandford, Narayan \& Romani (1986) we assume a Gaussian
distribution of ionized matter, keeping in mind that for IDV a much
smaller scale height $H\sim 100$pc is required than the ISM scale height
in the galactic disk.

We concentrate on the modulation index derived from the plateau of the SF at
large timelags and on the slope for shorter lags. It turns out that
in weak scintillation the slope of the SF is
either quadratic SF$(t) \propto t^\alpha$, $\alpha =2$ for steep turbulent
spectra $\beta \ge 4$ or $\alpha = \beta - 2$ for $\beta < 4$.   

In strong and quenched scintillation the SF shows a broken power law
below the maximum and this slope also depends on the power-law index $\beta$
of the turbulent spectrum $\Phi(q) = C_N^2 q^{-\beta}$ of density fluctuations
in the ISM. For the limiting case $\beta = 4$ discussed in Blandford, Narayan
\& Romani (1986),
this slope is $\alpha = 1$ in the quenched and strong case, while
for a Kolmogorov-like spectrum $\beta = 11/3$ the slope is
$\alpha = 5/6$ for strong scattering and $\alpha = 2/3$ for quenched
scintillation. For quenched scintillation, the slope is generally
$\alpha = \beta -3$. 
\begin{table}[t]
\begin{center}
\begin{tabular}{ccccccc}
\hline
 model & $H$ [pc]  & $\beta$ & $C^2_N$ [m$^{-\beta -3}$] & $\theta_{S6}$ [$\mu$as] &
 $\theta_{S20}$ [mas] & $v$ [km/s] \\
\hline
K     & 70 & 11/3  & $10^{-3}$ & 40 & 0.4 & 5 \\
S1    & 70 & 4.567  & $ 10^{-10.6}$ & 40 & 0.16 & 14 \\
S2    & 140 & 4.567  & $10^{-12.15}$ & 40 & 0.16 & 30 \\
\hline
\end{tabular}
\end{center}
\caption{Parameters for the structure function models. The result for
  $\lambda 6$cm and $\lambda 20$cm are shown in Fig.2 and Fig.3. The angular
  size $\theta_{S6}$ and $\theta_{S20}$ are the source size at $\lambda 6$cm
 and $\lambda 20$cm. }
\label{tab1}
\end{table}

We have computed theoretical structure functions dominated by quenched
scintillation and compared them to the observed data taken at $\lambda 20$cm
and $\lambda 6$cm (Fig.2 and Fig.3). The parameters are the power-law index
$\beta$, the strength of density fluctuations $C_N^2$, the scale height $H$
and the velocity of the observer relative to the ISM cloud. From simultaneous
fitting of the SF's at 6\,cm and 20\,cm we can find plausible parameters
and the size of source at these wavelengths. The model parameters are
summarised in Table 1. It is clearly visible from Fig.2 and Fig.3 that a steep
spectrum is preferred.

 \begin{figure}[t]
 \begin{center}
 \psfig{file=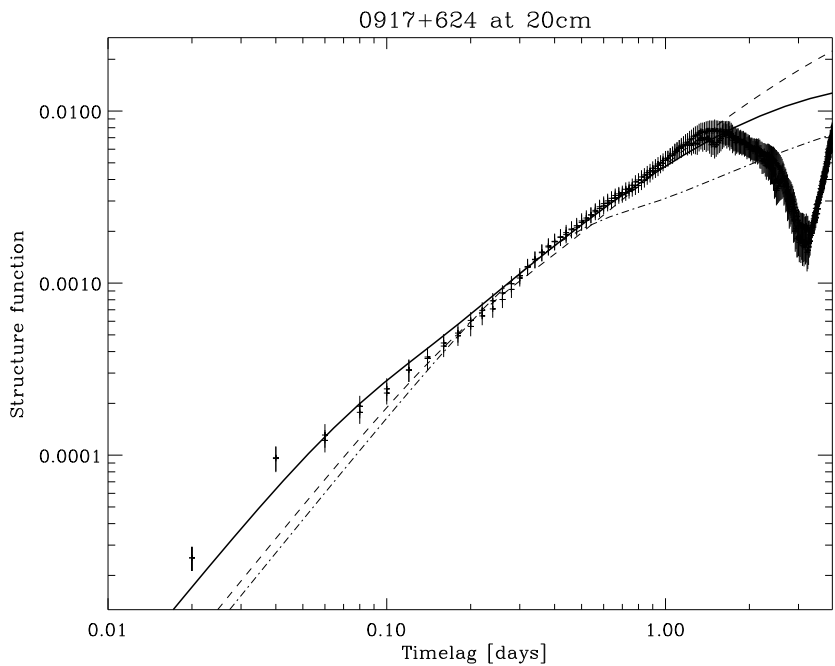,height=9cm}
 \caption{Structure function of total flux variations of 0917+624 at 20\,cm
          in 1989. The data have recently been analysed by Qian et al.
          (2001). Over-plotted on the SF derived from the measurement are
          models for quenched scintillation according to Table 1. The
          dash-dotted line shows the model with a Kolmogorov spectrum (K).
          The steep spectra models
          (S1 - dashed) and (S2 - solid) give better fits for the 20cm data.}
 \end{center}
 \end{figure}

 \begin{figure}[t]
 \begin{center}
 \psfig{file=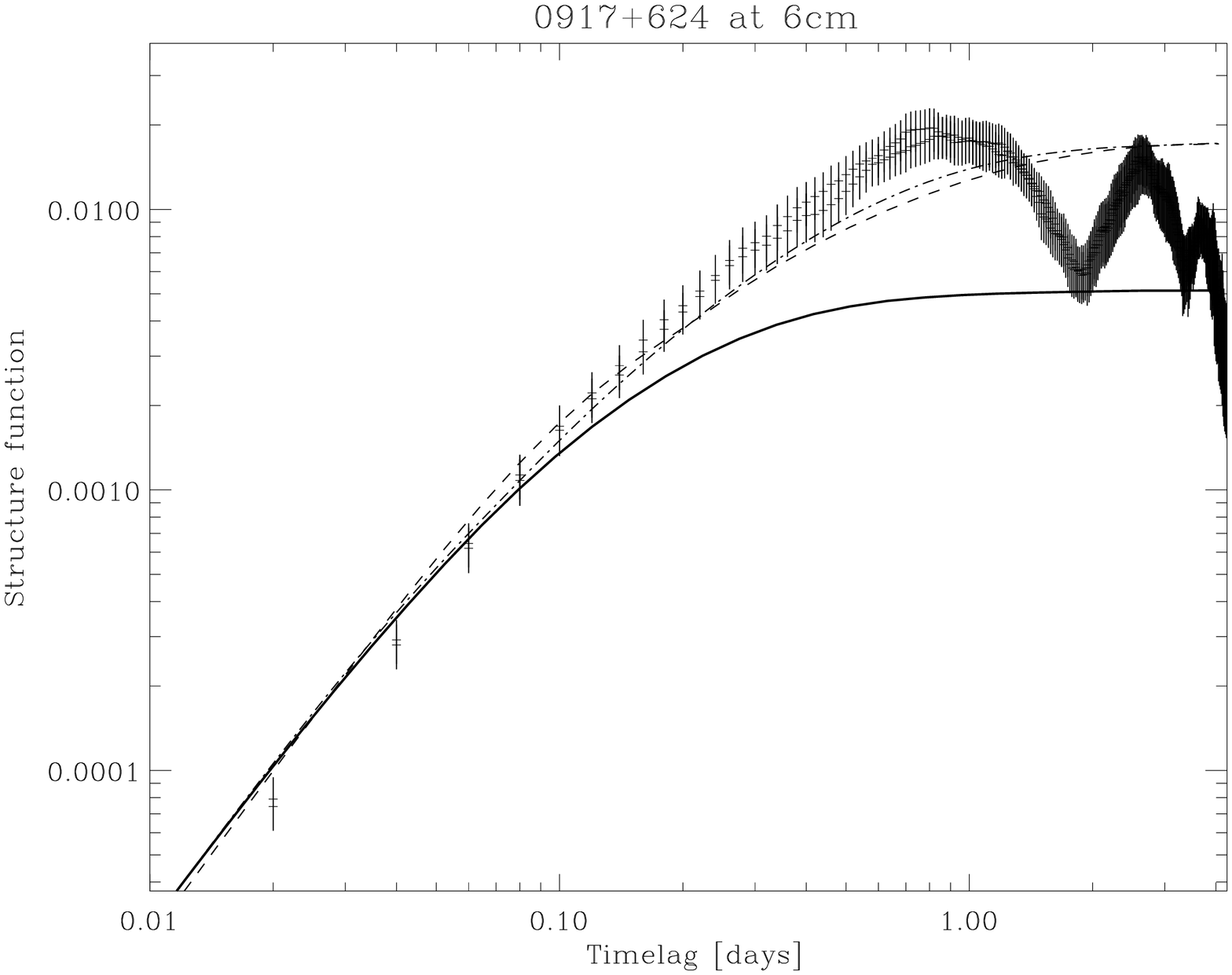,height=9cm}
 \caption{Structure function of total flux variations of 0917+624 at 6cm
          in 1997. The data come from one of the best sampled light curve
          at this wavelength. Over-plotted on the SF derived from the
          measurement are the same
          models for quenched scintillation like in Fig.2 applied to
          $\lambda 6$cm. }
 \end{center}
 \end{figure}

In particular the slope $\alpha \approx 1.3$ at $\lambda 20$cm indicates
a steep spectrum $\beta = 4.3$, when only quenched scintillation is assumed.
The turbulent spectrum has to be even steeper because the SF flattens
before turning over into the $\alpha = \beta -2$ slope dictated by the
Fresnel cut-off at smaller timelags. Together with the weak curvature
near the plateau for large timelags, this points to the slope
$\beta \approx 4.6$ of the models S1 and S2.

\section{Conclusions}
Based on IDV observations of several Quasar and BL Lac radio cores with
apparent brightness temperatures in the range of $10^{16}$--$10^{21}$\,K we
investigated the possibility of intrinsic variations due to the propagation
of thin relativistic shocks. We find that no source with $T_b > 10^{16}$\,K
can be explained by that model without calling for Doppler factors
larger than 20 and strong alignment between the jet and the line of sight.
The required
alignment firmly rules out this hypothesis as an explanation for all
IDV sources.

Based on refractive interstellar scintillation IDV can be explained by
turbulence in the ISM. The scale size of the ionized gas responsible
for IDV is about 100\,pc in the case of 0917+624, and scintillation is
quenched by the source size, which is larger than the refractive or Fresnel
scale in the ISM. A fit to the structure functions at
$\lambda 20$cm and $\lambda 6$cm indicates a steep power law
($\beta \approx 4.6$) for $\Phi(q)$ fluctuations. This corresponds to an energy
spectrum $\sim q^{-2.6}$ that is much steeper than both a Kolmogorov spectrum
and the $q^{-2}$ spectrum for compressible turbulence, 
but a bit shallower than $\sim q^{-3}$ expected for two-dimensional
turbulence. The length scale probed in the ISM by these measurements
are between $5\,10^8$\,m and $2\,10^9$\,m at the peak of the
structure functions. The slope of the turbulent spectrum is derived
from structure functions at small timelags and the corresponding spectrum
extends an order of magnitude down to smaller spatial scales.
Compressible turbulence is not unexpected at these scales, if turbulence
is driven by shocks from supernovae or by stellar winds.

Scintillation in 0917+624 is quenched by the source size, which is one
parameter of the theoretical fits to the SFs. The required sizes are
$\sim 40 \mu$as and $\sim 0.4$\,mas at $\lambda 6$cm and $\lambda 20$cm
respectively.
At $\lambda 6$cm, the flux of 0917+624 (redshift $z=1.446$) is 1.5Jy.
Combined with the derived source size this implies $T_b = 10^{14}$K.
Again Doppler factors of about 100 are needed to avoid the IC catastrophe
or Doppler factors of 1000 to arrive at equipartition temperature.
The sizes derived for 0917+624 from structure function models might be
changed, if a degeneracy in the model parameters exists. The most plausible
direction is an even closer ISM screen with a higher level of turbulence.
In this case the angular Fresnel scale gets larger and larger source
sizes are allowed.

%
%






\section*{References}
\reference Begelman, M.C., Rees, M.J., \and Sikora, M. 1994, ApJ, 429, L57
\reference Blandford, R., Narayan, R., \and Romani, R.W. 1986, ApJ, 301, L53
\reference Coles, W.A., Frehlich, R.G., Rickett, B.J., \and
  Codona, J. L. 1987, ApJ, 315, 666 
\reference Heeschen D.S., 1984, AJ 89, 1111
\reference Heeschen D.S., Krichbaum T.P., Schalinski C.J., Witzel A.,
   1987, AJ 94, 1493
\reference Jones T.W., O'Dell S.L., 1977, ApJ 214, 522
\reference Qian S.J., Quirrenbach A., Witzel A., et al. 1991,
   A\&A, 241, 15 
\reference Quirrenbach A., Witzel A., Krichbaum T.P., et al., 1992,
   A\&A 258, 279
\reference Readhead, A.C.S. 1994, ApJ, 426, 51
\reference Rickett, B.J., 1990, ARA\&A, 28, 561
\reference Rickett, B.J., Quirrenbach, A., Wegner, R.,
   Krichbaum, T.P., \and Witzel, A. 1995, A\&A, 293, 479
\reference Simonetti J.H., Cordes J.M. \and Heeschen D.S., 1985, ApJ 296, 46
\reference Spada, M., Salvati, M., \and Pacini, F. 1999, ApJ, 511, 136
\reference Wagner S.J., Witzel A., 1995, ARA\&A 33, 163 







\end{document}